# Testing Security Policies for Distributed Systems: Vehicular Networks as a Case Study

Mohamed H.E. AOUADI, Khalifa Toumi, Ana Cavalli
IT/ TELECOM & Management SudParis, EVRY, F-91011
{Mohamed.Aouadi, khalifa.toumi, Ana.Cavalli}@telecom-sudparis.eu

*Abstract*—Due to the increasing complexity of distributed systems, security testing is becoming increasingly critical in insuring reliability of such systems in relation to their security requirements. . To challenge this issue, we rely in this paper[1] on model based active testing. In this paper we propose a framework to specify security policies and test their implementation. Our framework makes it possible to automatically generate test sequences, in order to validate the conformance of a security policy. This framework contains several new methods to ease the test case generation. To demonstrate the reliability of our framework, we present a Vehicular Networks System as an ongoing case study.

## I. INTRODUCTION

It is increasingly difficult to ensure the respect of applications to their security requirements. This difficulty is due to the such systems' complexity level of, variety, and increasing distribution as well as the high degree of reliability required for their global functioning. To guarantee such a respect, we need to generate exhaustive test suites including all possible scenarios. To reach this test integrity, we rely in this paper on model-based methods. These methods require a formal specification of the system (a model). To meet security requirements, we specify a security policy in XACML (eXtensible Access Control Markup Language. This security policy needs to be integrated into the initial model. To do so we develop an approach to automatic integration of security rules into any initial model. Once we have a secured model we perform tests to verify the conformance of the model to its security policy. These tests are generated by TestGen-IF, a testing tool developed in our laboratory. To gain time and effectiveness when using our tool, we propose an approach that derives test cases automatically. We also develop a Graphical User Interface (GUI) to facilitate its use. To demonstrate the reliability of our framework, we carry on a case study which is a service of the Vehicular Networks. Vehicular networking serves as one of the most important enabling technologies required to implement a myriad of applications related to vehicles, vehicular traffic, drivers, passengers and pedestrians. These applications are more than novelties and far-fetched goals of a group of researchers and companies [1], [2], [3]. Intelligent Transportation Systems (ITS) that aim to streamline the operation of vehicles, manage vehicular traffic and assist drivers with safety and other information by many services. An example of such services is Dynamic Route Planning (DRP) in which the driver receives an optimal route to reach his destination. This route is calculated dynamically by the ITS control center and takes into account the environment conditions (traffic, congestion, weather, accidents, etc.).

In this paper, we propose an approach that makes it possible to validate security rules. Our approach manipulates three different inputs: a functional specification of the system based on a well-known mathematically based formalism, the Extended Finite State Machine (EFSM, see section III); a specification of the security policy (based on XACML [4]) that we wish to apply to this system; and an implementation of the system. Our solution provides a new specification of the system that takes into account the security policy (we call it secure functional specification), then it will generate tests to check whether the implementation of the system conforms with the secure functional specification. This approach will be enriched by some improvements in the TestGen-IF tool, namely a method to generate test purposes and a graphical user interface (GUI). The reliability of our approach will be demonstrated by a case study of the Vehicular Networks.

The main contributions of this paper are as follows:

- The specification and implementation of an approach to integrate XACML security rules into a functional model described by an EFSM.

- We propose a new method to semi-automatically derive test scenarios. This method, based on TestGen-IF simulator, is illustrated by a testing scenario.

- We provide a method to automatically derive testing scenarios directly from the formal model as test cases. These scenarios allow checking relevant security and interoperability properties of an implementation under test. We use the TestGen-IF tool to automatically obtain the set of tests. Real scenarios of vehicular networks communications are used in order to highlight the advantages and the new functionalities of our approach.

- we propose some improvements in the TestGen-IF tool in the form of a method to automatically generate test purposes and a GUI.

The rest of this paper is organized as follows. In section II we discuss related work. Section III defines the two concepts of XACML and EFSM. In section IV, we explain the approach to integrating these security rules with an existing specification in EFSM as well as related algorithms. In section V, we present a case study, a DRP service with security features, a complementary approach to generating test objective, the GUI that we developed, and the results through the generated test cases for the validation of security rules and the experimental result. Finally, section VI presents our conclusions and some directions for future work.

---

[1]This work is supported by the Inter-Trust project.

## II. RELATED WORK

Most previous work has focused either on the description of the policy itself or on the verification of rules. Security rules are defined with modalities (such as permission, prohibition, and obligation) that express possible constraints on the behavior of the system.

In [5] the authors propose a testing strategy for automatically deriving test requests from an XACML policy and describe their pilot experience in test automation using this strategy. In [6] the authors give an overview of existing security testing approaches and, based on that, develop a novel classification for model-based security tests along the two dimensional-risk and automated test generation. In other works [7], [8], the authors propose approaches based on active testing. In [7], the authors propose a framework that specifies security policies and tests their implementation and the behavior of the system is described using the EFSM formalism [9]. In [10] the authors propose a framework to specify security policies and test their implementation on a system. This framework is based on a specification of the system in EFSM formalism and on a specification of the security policy based on the OrBAC model.

Our approach distinguishes itself from these propositions by assumptions on the policy and the method used to generate test sequences. First, we make no assumption about the description language of the policy. Instead, we propose a framework to specify rules in XACML so that we can apply them to our mathematical model. XACML supports Attribute-Based Access Control (ABAC) and can implement an access control model based on RBAC or OrBAC models, making our study more general than [10] where authors use OrBAC and than [11] where authors use O2O. Then, we generate a whole set of test cases automatically using a different test generation tool developed in our laboratory. This latter is well adapted to the EFSM formalism which makes our approach different from [10] and [8].

## III. PRELIMINARIES

### A. The EFSM formalism

In order to model the initial system as well as the security policy, we choose to use the EFSM formalism. This formal description is used not only to represent the control portion of a system but also to properly model the data portion, associated variables, and the constraints that affect them.

An EFSM is an augmentation of the ordinary Finite-State Machine [12] with guard functions (predicates) and action functions. We consider that a transition can be executed only when an input is received and the predicate is true.

*Definition 1:* An extended finite-state machine is a 5-tuple

$$E = \langle Q, \vec{v}, \Sigma, T, q_0 \rangle$$

where:

1) $Q$ is a finite set of states and $q_0$ is the initial state;
2) $\vec{v} = (v_1, ..., v_n)$ is a vector of typed variables;
3) $\Sigma \subseteq (L_i \times L_0)$ is a nonempty set of input/output alphabet, where $L_i$ and $L_0$ are the set of inputs and outputs, respectively;
4) $T$ is a set of transitions, defined by a tuple $\langle q, \sigma, \wp, a, q^0 \rangle$ where,
   a) $q$ and $q^0$ are the source and the target state, respectively;
   b) $\sigma \in \Sigma$ is the input/output of the transition;
   c) $a$ is an action of a transition. Variables values can be updated after transition execution;
   d) $\wp$ is a predicate over $\vec{v}$.

*Definition 2:* The characteristic function of a subset $A$ of a set $C$ is a function

$$\chi_A : C \to \{0, 1\}$$

defined as

$$\chi_A(x) = \begin{cases} 1 & \text{if } x \in A \\ 0 & \text{otherwise} \end{cases}$$

We illustrate the notion of EFSM through a simple example described in Figure1. The ESFM shown in Figure 1 is composed of two states $S_0$, $S_1$ and three transitions labeled with two inputs $A$ and $B$, two outputs $X$ and $Y$, one predicate $P$, and three tasks $T$, $T^0$ and $T^{00}$. The EFSM operates as follows: Starting from state $S_0$, when input $A$ occurs, the predicate $P$ is tested. If the condition holds, the machine performs task $T$, triggers output $X$, and passes to state $S_1$. If $P$ is not satisfied, the same output $X$ is triggered, but action $T^0$ is performed and the state loops over itself. Once the machine is in state $S_1$, it can come back to state $S_0$ if it receives input $B$. If so, task $T^{00}$ is performed, and output $Y$ is triggered.

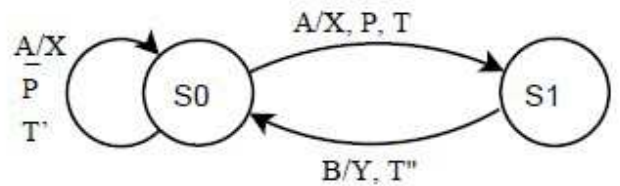

Fig. 1. Example of simple EFSM with two states

### B. XACML

XACML [4] is a platform-independent extensible markup language (XML)-based language for the specification of access control policies. An XACML policy consists of a target, a set of rules, and a rule-combining algorithm. The target specifies the subjects, resources, actions, and environments to which a policy can be applied. If a request satisfies the target of the policy, then the set of rules of the policy is checked, or else the policy is skipped. The rule is composed by a target, which specifies the constraints of the requests to which the rule is applicable. The rule has a condition, which is a Boolean function evaluated when the rule is applicable to a request. If the condition is evaluated as true, the result of the rule evaluation is the rule effect ("Permit" or "Deny"); otherwise a "NotApplicable" result is given. If an error occurs during the application of a policy to the request, "Indeterminate" is returned. Each policy has at least one rule (possibly more). There must be at least one rule in a policy that matches the incoming request so that the policy can be deemed applicable to that request. The Sun XACML engine determines whether a rule is applicable to an incoming request by evaluating the target and optional condition (if it exists).

Target: A policy can have multiple rules. But it is not necessary to evaluate all such rules for a given request. A rule has a target element similar to the policy's target element. The role of this target element is to decide whether a rule should be evaluated for a given request. If no target exists, the rule is evaluated for all requests applicable to the policy.

Condition: You can treat the condition as the core element of a rule. Within the condition we specify the exact authorization logic, which always contains a Boolean expression. Based on the outcome of the Boolean expression, the rule is evaluated as true or false. We can use certain functions within the condition element to implement authorization logic.

Figure 2 shows the structure of a policy and its rule.

We choose to use XACML for several reasons. First, XACML is a standard ratified by standards organization OASIS and a policy language implemented using XML. Second, XACML supports Attribute-Based Access Control (ABAC) and can implement an access control model based on RBAC or OrBAC models, making our study more general than the works cited in sectionII. Finally, XACML provides fine-grained authorization with a high level of abstraction by means of policies and rules, making our study more exhaustive by providing more flexible security rules.

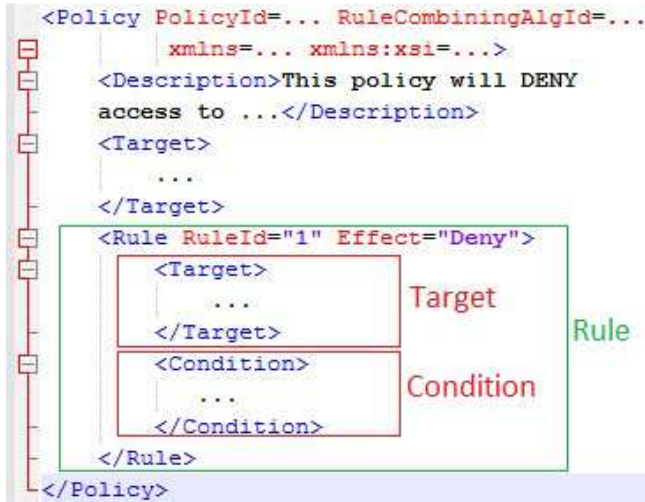

Fig. 2. The main components of an XACML policy

## IV. INTEGRATION METHODOLOGY

The initial specification does not consider security issues and does not deal with security rules. This specification is provided in the form of an EFSM. We need to consider security requirements and integrate them. To do so, we integrate them into the initial specification using a specific algorithm. This algorithm automatically integrates the security rules into the initial specification in the form of an EFSM. Our approach includes three steps. First, the algorithm seeks for the rules to be applied to each transition of the specification and derives a simple automaton from this set of rules. Then, it integrates the automaton with the initial specification. At the end of the process, this integration generates a new specification that takes into account the security requirements. While the main effects of an XACML rule is Permit or Deny, our approach contains two algorithms. The first one deals with the permissions and the second one deals with the prohibitions.

### A. Permission integration

This process of the algorithm begins by seeking for the transitions with a permission rule to integrate. Once this transition is found, the algorithm can either create a new predicate or strengthen an existing predicate depending on the existence or the absence of a predicate in that transition. The strengthening or creation of a predicate is done by adding the conditions in the XACML security rule. This algorithm is described below.

Algorithm 1: Permissions integration

---

Require: The transition $Tr$ that maps the permissions.
Each $permission_i$ applies to a $condition_i$
1: if ( $\exists$ associated predicate $P$ ) then
2:    $P := P \wedge ( \vee_i (condition_i) )$
3: else
4:    create predicate $P := \vee_i (condition_i)$
5: end if

---

Figure 3 gives an example. In the left transition, the system can pass from S1 to S2 when it receives input $A$. If the permission involves a condition $C$, the transition is modified by creating a predicate, as in the left transition. This predicate returns "true" if the condition is satisfied.

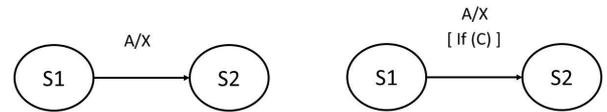

Fig. 3. Permission integration

### B. Prohibition integration

This process of the algorithm begins by seeking for the transitions with a prohibition rule to integrate. Once this transition is found, the algorithm can either create a new predicate or strengthen an existing predicate depending on the existence or the absence of a predicate in that transition. The strengthening or creation of a predicate is done by adding the opposite of each condition in the XACML security rule. This algorithm is described below.
Algorithm 2: Prohibition integration

---

Require: The transition $Tr$ that maps the prohibitions.
1: if ( $\exists$ associated predicate $P$ ) then
2:    $P := P \wedge ( \vee_i (\neg condition_i) )$
3: else
4:    create predicate $P := \vee_i (\neg condition_i)$
5: end if

---

An example is shown in the Figure 4. In the left transition, the system can pass from S1 to S2 when it receives input $A$. If the rule specifies that the system is prohibited from sending output $X$ in condition $C$, the transition is modified by the creation of a corresponding predicate, as in the left transition.

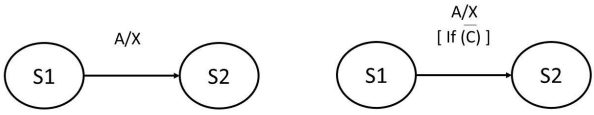

Fig. 4. Prohibition integration

## V. CASE STUDY

### A. DRP description

In order to demonstrate the reliability of our approach, we conduct a case study. Supported by the INTER-TRUST project, this case study is an application to a service called Dynamic Route Planning (DRP). The DRP service aims at providing the driver with an optimal route to reach his destination. This optimal route must take into account different changing factors, such as traffic, the weather, and the state of the road traffic. This route also allows for a reduction in travel times by means of choosing the most efficient succession of roads. The user (driver) wants to reach his destination by the optimal route. Therefore, he must activate the service through his client interface. Then, he must activate only the DRP service. Once the service is activated, the system (modeled as the control center or the server) must check if the user is authorized to access the service. Here we consider at first a simple service with various features, such as those commonly used in GPS devices. First of all, the service is open to anyone. Without using the security policy, any user can access the service with premium-user privileges. Such access causes different security problems, such as denial of service and malicious use of shared data.

To tackle this problem, we specify a security rule that protects the information within the organization by preventing illegitimate users from using the DRP service. For this purpose, the security rules restrict access to the service to authorized users. To do so, the user has to introduce a valid username and password. Another issue is that security rules restrict parts of the service to premium users only. For example, a regular user cannot get a route to a destination outside France. Therefore, the system (the server) checks the GPS position before calculating the route: If the user is a premium user, then the service allows him to have an international route. However, if the user is a regular user he is allowed to get only a national route. We model the specification of this system by the EFSM shown in Figure 5. This EFSM models the internal behavior of the system (server) without security properties. This EFSM has three states S1, S2, and S3. S1 presents the initial state of the server before any interaction with the user. S2 shows that the server is connected to the user and waiting for the desired destination from the user. S3 shows that the service is monitoring the navigation of the user, indicating that it already calculated the optimal route and is just waiting for any other request from the user to stop the service or to calculate another route.

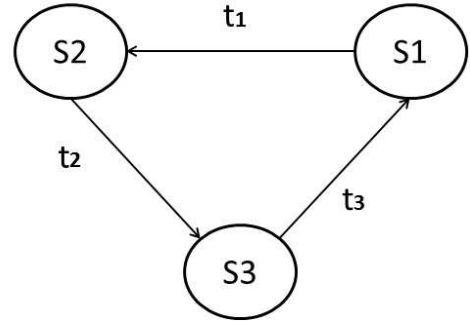

Fig. 5. Initial model of the server

The transitions marked $t_1$, $t_2$, and $t_3$ are defines as follows:

- $t_1$: ask_access(login,password,GPSposition) / access_authorised
- $t_2$: ask_for_route(destination,class) / response(optimalRoute)
- $t_3$: exit_service / exit_ok

In this section, we aim to illustrate the integration methodology presented in section IV and present some complementary approaches and methods. The integration methodology will be illustrated by the scenario presented in subsection V-A and will be illustrated in subsections V-C and V-D. In subsection V-E, another scenario will be presented and will be used to introduce a semi-automatic method to generating test cases (in subsection V-G) and an approach to automatic generating of test objectives (in subsection V-I). In subsection V-J we present our new GUI. Subsection V-K presents the experimental results and subsection V-L discusses them.

### B. TestGen-IF tool

The tool was developed by our research team at Telecom SudParis [13], [14] for modeling and simulating asynchronous timed-systems such as telecommunication protocols or distributed applications. It is based on active testing techniques, allowing automatic generation of test-cases from a formal description of the studied system. The generation is made according to specific objectives called test purposes.
TestGen-IF tool allows construction of the accessiblity graph from an IF specification. Therefore, the inputs necessary for TestGen-IF tool are the formal functional specification of the system and the specification of test objectives that we wish to check on the system implementation. Then, the tool makes a partial exploration of the states space of a model, guided by test objectives. The automatic test generation module, written in C++ code, implements an automated test generation algorithm called Hit-or-Jump [15].

### C. Security policy specification

In this case study, we implement some security properties to get a secure system that considers security issues. In the security policy, services must be accessed only by authorized users. To access the DRP service, the user must introduce a valid login and password and must have a valid GPS position. Once connected to the service, the regular user can use only basic functionalities. For example, the user cannot use international navigation, which is reserved for premium users. This security policy is described in XACML. Each security rule contains one or more conditions. We can summarize the conditions as follows:

- $C_1$ - GPS position is in France
- $C_2$ - login and password are valid
- $C_3$ - the user's class is PREMIUM
- $C_4$ - the user's class is REGULAR
- $C_5$ - GPS position is valid
- $C_6$ - GPS position is not in France
- $C_7$ - destination is not in France
- $C_8$ - destination is in France

Thus, the security policy contains the three following rules:

- Rule 1: The server grants access to a user if he has a valid couple (login,password) and a valid GPS position.
- Rule 2: The server gives an international optimal route for PREMIUM users.
- Rule 3: The server does not give an international optimal route for REGULAR users.

### D. Security policy integration

By applying our algorithm to the initial model and the security policy described in subsection V-C we obtain a secured model of the system that takes into account the security policy. This model is presented by the EFSM shown in Figure 6.
The final model contains the same states, but some transitions are modified and other transitions are added. We have two new transitions:

- $t_4$:  ask_access (login ,password, GPSposition) / access_denied
- $t_5$:  ask_for_route(destination, class) / need_premium_class

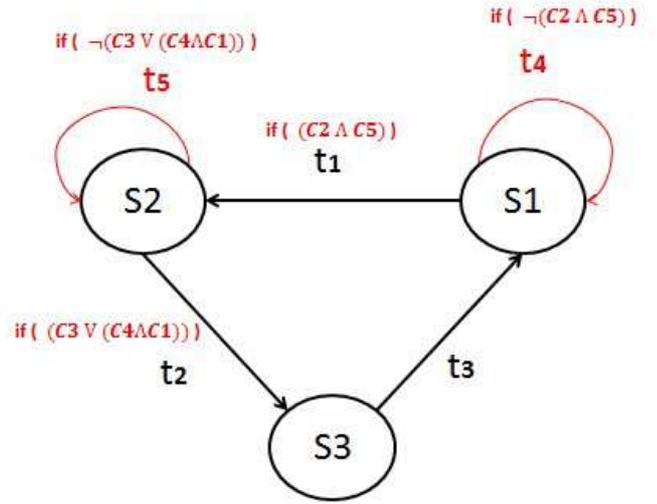

Fig. 6. Final (secure) model of the server

| Model | Number of states | Number of transitions | Number of signals |
|---|---|---|---|
| Initial model | 3 | 3 | 6 |
| Final model | 3 | 5 | 8 |

TABLE I. INITIAL MODEL OF THE SERVER

Moreover, transitions are modified, and two conditions are added to $t_1$ and $t_2$. Table I compares the model before and after security policy integration.

### E. Vehicle to Infrastructure (V2I) scenario with negotiation

Trust negotiation scenario:
To meet the secure interoperability requirements of the system we propose a trust negotiation process. This scenario meets the following requirement:

"Before exchanging personal information the vehicle must trust the control center"

The scenario starts when the user asks to activate a service. Once the DRP service is activated, the control center checks if the user is authorized to access the service. To have access, the vehicle must be located in the coverage area of the control center so it needs its current position. It also needs his login and password to do the identification and authentication. However, due to security reasons, the vehicle cannot send personal information anyhow to anyone. Therefore, the vehicle must trust the control center. This trust can be established after a trust negotiation between the user/vehicle and the control center. In our model this negotiation is done by the exchange of security certificates. The vehicle automatically asks the control center for a certificate. It can accept it, reject it, or delegate the user to make a decision. If a certificate is accepted then it will move automatically to the next state (cf. EFSM Figure7) to accomplish the identification and authentication. If the certificate is rejected the control center will try again with another certificate. For the third case, the user will make a decision instead of the vehicle.

Formalization of the scenario:

The Figure 7 illustrates the EFSM that describes the internal behavior of the vehicle with the negotiation process. This EFSM has the following states:

- Off-line: is the initial state of the vehicle. We consider that any vehicle at this state is not connected to the control center and is not logged onto the service. Once the user activates the DRP service by the message **activate_service**, the vehicle asks the control center to activate the DRP service (by the **request_service** message) and moves to the state Wait.

- Wait: The service is activated and the vehicle is waiting for the control center to reply. The control center replies and asks for information (login, password, position, identity, etc.) by sending the **request_information** message to the vehicle. The vehicle automatically answers by asking the control center for a certificate (by the **request_certificate** message) which makes it move to the state wait_certificate.

- Wait_certificate: The vehicle is waiting for the certificate. Depending on the value of the received certificate, the vehicle disagrees(output **disagree_certificate**), agrees( output **agree_certificate**), or delegates the user who will decide. Then, the system will move to the state wait_decision or wait_info, or stay in wait_certificate state depending on its answer.

- Wait_decision: The vehicle is waiting for the decision of the user. The user can accept the terms of the certificate and send an **agree** message. Then, the vehicle asks him for some information to log into the service (output **require_info_login**) and moves to the state wait_info. The user can also disagree with the certificate then the vehicle will ask the control center for another certificate (output **request_certificate**). Therefore, the vehicle moves back to the state wait_certificate.

- Wait_info: The vehicle is waiting for necessary information from the user to log in. Once the user gives this information (input **give_info** with login and password variables), the vehicle forwards the information to the control center (output **response_info**) and moves to the **wait_access** state.

- Wait_access: The vehicle is waiting for the response of the control center which will give it access or not to the service (input **access_ok** or **access_denied**). Once the control center accepts the access, the vehicle will move to the state logged_in. If the access is denied, the vehicle moves back to wait_info state.

F. IF Specification

The description of a system in IF [16], [17] consists of the definition of data types, constants, shared variables, communication signals, and processes. In order to use the TestGen-IF tool, we transcribed the formal model into an IF model. In any IF model we must have a system and some processes. In our model the system is V2I. The processes are the actors of the scenario, so they can be the vehicle, the user, and the control center. To simplify our model we will consider only the vehicle process. The other actors are considered as the environment. In Figure 8 we represent a short code of our system in IF language.

```
system V2I;
/*constant definition*/
const SD = 100;
/*const NbService = 10;
...
/*type definition*/
type states = enum off_line, wait, wait_certificate,
        wait_info, wait_access, logged_in
    endenum;
....
/* Signals definitions */
/* channel 1 --- vehicle to CC */
signal response_info(login, password,position) ;
signal request_service(service);
....
/* Main process */
 process vehicle(1);
 /*local variables*/
var servicex service private;
...
 /* States specification */
state wait ;
    input GiveInfoLogIn(loginx,passx);
    output Forward_Inf_log_in(loginx,passx) ;
    nextstate waitForLogIn ;
.....
endstate;
....
endprocess;
endsystem;
```

Fig. 8. A sample code of the V2I under DRP system specification in IF

G. A semi-automatic method to get all the test scenarios

TestGen-IF tool allows an interactive simulation of the system specified. The instances of the processes described in the specification are shown with corresponding parameters, as well as all the transitions possible from current state. Therefore, all scenarios can be deducted from the interactive simulation. Figures 9 and 10 illustrate the simulation of the system specification we implemented. Usually, the simulation is used to verify the specification and test all the transitions and behavior of the system according to some data parameters. We propose a new use for the interactive simulation, namely to know the number of possible tests and distinguish them.

This method is semi-automatic because the simulator shows all the possible transitions for the current state, the user has to choose manually one transition, and the simulator calculates the next state and then shows the expected output and the expected next state. Each sequence of transitions from an initial state to a final state constitutes a test scenario. The tester does not have to know all transitions, all inputs, nor all possible values of a variable anymore. This effectively simplifies the test derivation in the case of a big system.

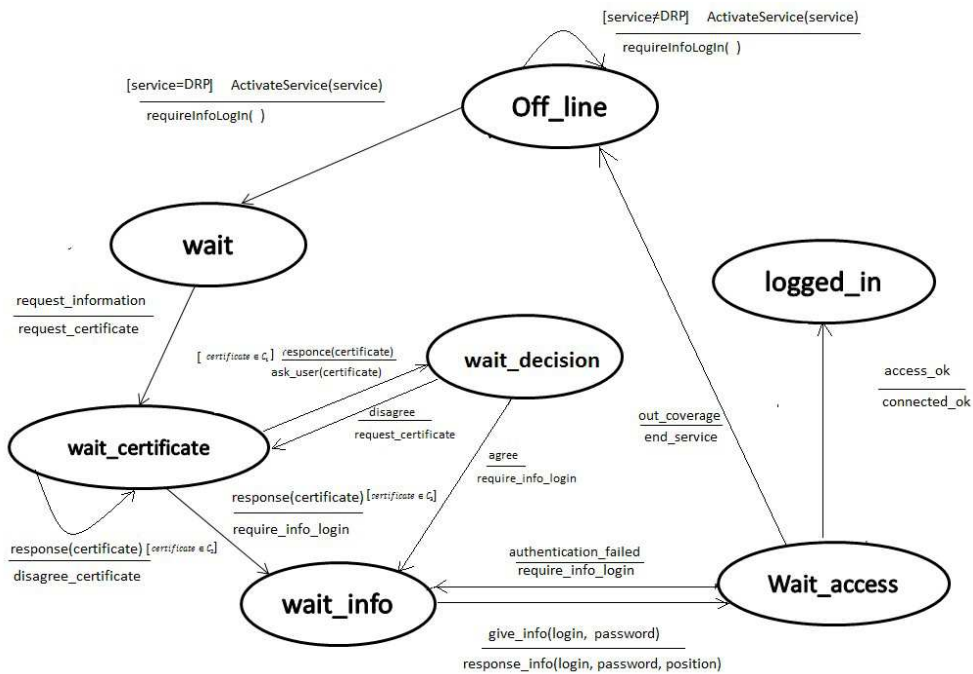

Fig. 7. EFSM of the vehicle with trust negotiation

Fig. 9. Initial phase of interactive simulation

Fig. 10. Interactive simulation after 2 steps

Moreover, this method needs only the IF specification and don't need test objectives.

### H. A formal approach to generating test objectives

Motivation

We consider the scenario of subsectionV-E. As seen in this subsection, to verify a security rule we need a set of test objectives. This set of test objectives is specified manually. This specification is possible in this case. However, it becomes time consuming with the risk of errors if we have a large number of possible parameter values. For instance, if we have infinite possible values (or a very large number of possible values) of the **certificate** parameter the test objectives specification becomes impossible (or time consuming). To tackle this problem we propose a formal approach that permits the automatic generation of a set of test purposes whenever a parameter has many possible values.

### I. The approach

The approach is based on an algorithm which takes the signal (input or output), the initial state, and the parameter as inputs. Based on the specification of the system provided by the EFSM, the algorithm will generate a set of test objectives.

Algorithm 1: Test Objectives Generation

Require: The initial state $S_1$, the signal input, the process process, and the variable parameter.

1: for each value$_i$ of the parameter output$_i$
   = g(value$_i$, S1, input) destination$_i$ = f
   (value$_i$, S1, input, output$_i$)

2: function f(p,s,i,o){
      return the state S2 destination of the transition
      ?i(p)!o which leaves from s }

3: function g(p,s,i){
      return the output o sent by the system when we
      apply i(p) on the state s }

4: write
   obj$_i$ = cond1 ∧ cond3 ∧ cond4 ∧ cond5
   cond1 = process: instance = {process}0
   cond2 = state: source: S1
   cond3 = state: destination: destination$_i$
   cond4 = action: input input(value$_i$)
   cond5 = action: output output$_i$

Example

Consider the system shown in Figure 7 which is in a **wait_certificate** state. Under this state, the vehicle is waiting for the certificate from the control center. The control center sends a certificate via the message (input to the vehicle) **response(certificate)**. Depending on the certificate value the system can go to the state **wait_decision**, **wait_info**, or stay in its current state. Our algorithm takes as input the current state **wait_certificate**, the input without parameter **response(certificate)**, and the parameter **certificate**. Then the algorithm defines all possible test objectives by calculating the output and the destination state for each test objective. By applying this algorithm to our case, study we obtain the three test objectives shown below.

obj$_1$ cond1 ∧ cond3 ∧ cond4 ∧ cond5
   cond1 = process: instance = {vehicle}0
   cond2 = state: source: wait certificate
   cond3 = state: destination: wait_info
   cond4 = action: input response(certificate01)
   cond5 = action: output require_info_login

obj$_2$ cond1 ∧ cond3 ∧ cond4 ∧ cond5
   cond1 = process: instance = {vehicle}0
   cond2 = state: source: wait certificate
   cond3 = state: destination: wait_decision
   cond4 = action: input response(certificate02)
   cond5 = action: output ask_user(certificate02)

obj$_3$ cond1 ∧ cond3 ∧ cond4 ∧ cond5
   cond1 = process: instance = {vehicle}0
   cond2 = state: source: wait certificate
   cond3 = state: destination: wait certificate
   cond4 = action: input response(certificate03)
   cond5 = action: output disagree_certificate

### J. TestGenIF interface

Another contribution of this paper is the specification and the development of a graphical user interface. This interface aims to simplify the tester tasks and to enhance the application efficiency. Figure 11 shows the interface of our system. It is composed of 3 parts:

- The first one contains the EFSM of the application. This one contains all the details about the transitions and the states.
- The second one offers the possibility to choose the test purposes. With this tool, the tester has only to choose the transitions to be tested. No commands or a configuration should be done.
- The third part will show the results that are the abstract test cases. Moreover, we have added a new extension of the TestGen-IF tool in the form of an execution engine. This engine will permit the test to be executed automatically. However, this process would have to be updated based on the application. Our work also resulted in the creation of a plugin that allows translation of the abstract test case into a concrete one for some web applications. However, this translation would need to be updated for use in other applications.

### K. Experimental results

*1) Fixing test objectives:* The test cases are generated by using an automatic test case generation tool, TestGen-IF [18], which is based on the Hit-Or-Jump [15] algorithm. The tool generates a test case guided by predefined test purposes which are sequences of conditions. A test purpose verifies that when the server communicates with its environment and if a security rule is satisfied, the server behaves as specified by the security rule. In order to validate the security between the server and its environment, we define for each security rule the corresponding test purpose and obtain seven test cases that cover all of the security policy. The first step in the experimentation is the specification of the system into an IF code format.

When translating the formal EFSM model into an IF specification, we consider the following variables' values:

- The couple (log$_i$, pwd$_j$)$_{i=j}$ is a valid couple (**login,password**).
- The **class** of the user is either **premium** or **regular**.
- We have two values of **GPSposition**: GPSin (valid GPS position) and GPSout (invalid GPS position).
- We have two values of **destination**: **destinationIn** (inside France) and **destinationOut** (outside France).

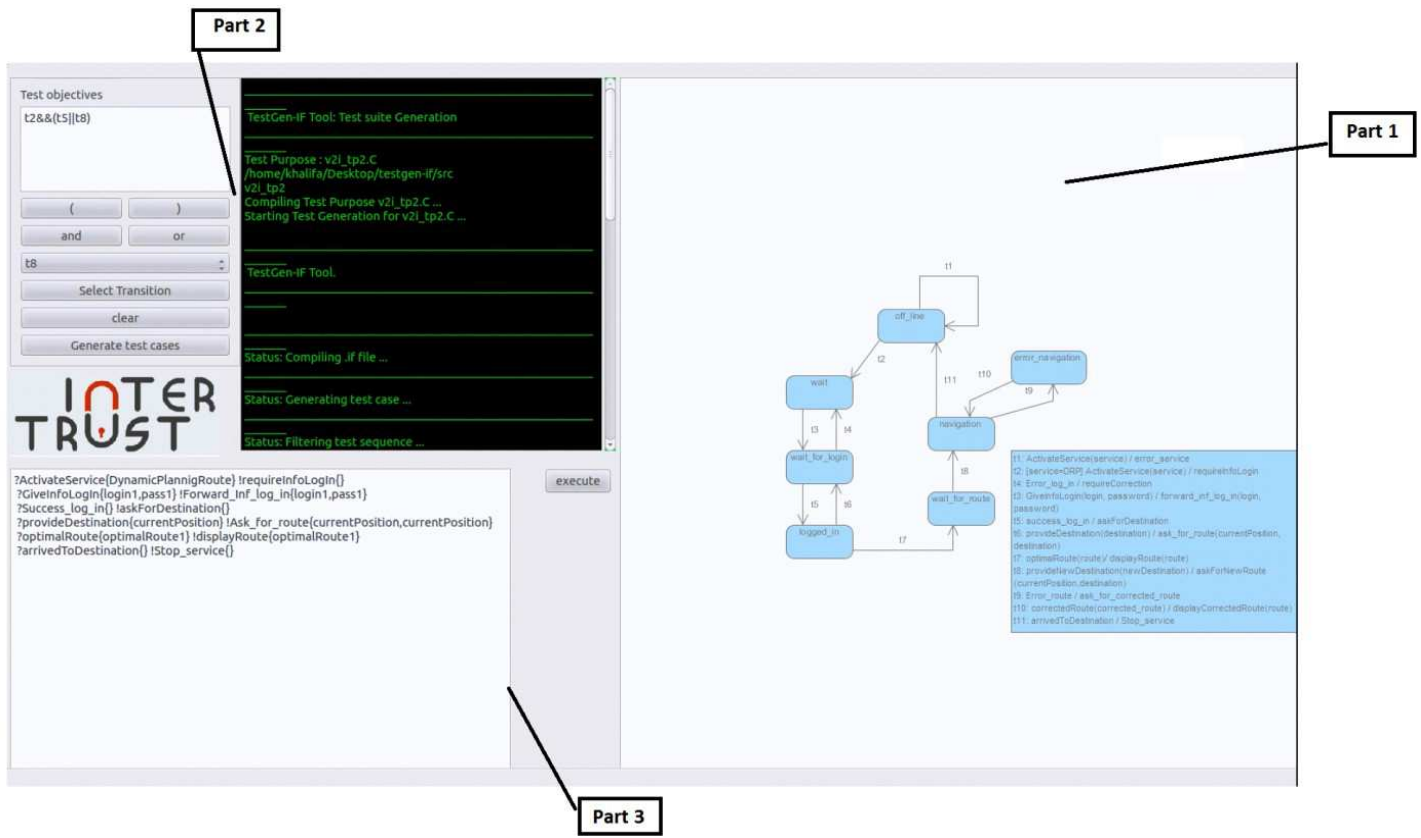

Fig. 11. The GUI interface of the TestGen-IF tool

While defining the test purposes, we notice that in some cases, we need more than one condition in a test purpose to verify a security rule. For example, when a user tries to access a DRP service in a foreign country, the server should check if the user's class is premium before granting an access. In this case, we define the test purpose as a sequence of two conditions. The first condition verifies that the user is in a foreign country. The second condition verifies that the user's class is premium.

2) Test case generation: Using the TestGen-IF tool we are able to generate a test sequence that verifies the two conditions. Figure 12 shows the test case generated for Rule 3 (presented in V-C). Note that the inputs and outputs are applied and observed by the tester. Another example is shown in Figure 13 which explains the test case generated for Rule 1.

```
?ask_access{log1,pwd1,GPSin} !access_autorised{}

?ask_for_route{destinationOut,regular} !response{optimalRoute}

?ask_for_route{destinationIn,regular} !response{optimalRoute}

?exit_service{} !exit_ok{}
```

Fig. 12. Test case related to the security Rule 3

```
//the tester introduces a wrong couple (login,password)
// and a not valid GPS position
?ask_access{log1,pwd2,GPSout} !access_denied{}

//the tester introduces a wrong couple (login,password)
// and a valid GPS position
?ask_access{log1,pwd2,GPSout} !access_denied{}

//the tester introduces a valid couple (login,password)
//and a not valid GPS position
?ask_access{log1,pwd1,GPSout} !access_denied{}

//the tester introduces a valid couple (login,password)
//and a valid GPS position
?ask_access{log1,pwd1,GPSin} !access_autorised{}

?ask_for_route{destinationOut,regular} !response{optimalRoute}

?ask_for_route{destinationIn,regular} !response{optimalRoute}

?exit_service{} !exit_ok{}
```

Fig. 13. Test case related to the security Rule 1

L. Discussion

The results allow the integration of security properties into an initial model with no security. The final (secured) model is different from the initial model because we have new transitions or conditions on the existing transitions. The number of modifications in the model depends on the number the complexity of security properties, making our approach flexible and adaptable to the XACML policy. Our work not only enables the securing of a model but also tests the security properties of this model to verify that the model respects the security properties. Our testing strategy is different from the methods cited in section II because our tool, TestGen-IF, offers high-performance test generation through its generation algorithm called Hit-or-jump [15]. This algorithm makes TestGen-IF faster than classical test generation tools (a gain of almost 20%) and less memory consuming. In addition, TestGen-IF avoids the state explosion and deadlock problems. Figure 14 shows the different steps to use our approach.

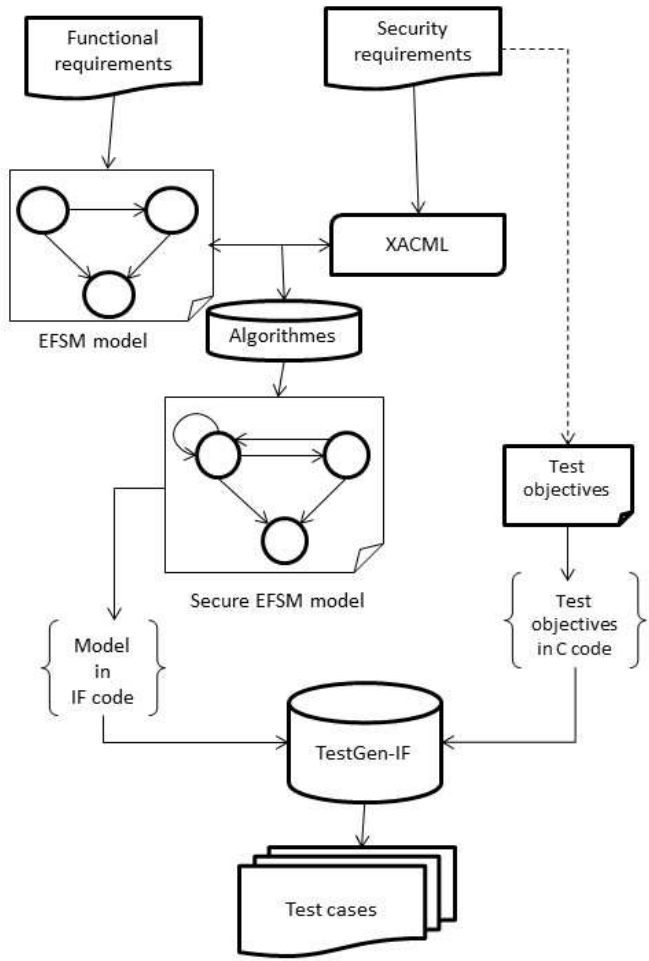

Fig. 14. The different steps to use the approach

## VI. CONCLUSIONS

In this paper, we present a framework that allows the testing of security policy of distributed systems. This framework is composed by several methods and approaches to best perform the specification, integration, and testing of such systems. We propose an approach that permits the automatic integration of a security rule in a system. Then we propose an approach to automatic generation of test objectives and we develop a GUI which improves the performance and facilitates the use of our testing tool called TestGen-IF. We plan to extend our work to consider timed security policies and to improve our approach to consider interoperability security policies.